# Nano-optical imaging of monolayer MoSe$_2$ using tip-enhanced photoluminescence

Chenwei Tang[1,2], Shuai Jia[3], Weibing Chen[3], Jun Lou[3], and Dmitri V. Voronine[1,4,5]

[1]Institute for Quantum Science and Engineering, Texas A&M University, College Station, TX 77843, USA

[2]School of Science, Xi'an Jiaotong University, Xi'an, Shaanxi 710049, China

[3]Department of Materials Science & Nano Engineering, Rice University, Houston, TX 77005, USA

[4]Department of Physics, Baylor University, Waco, TX 76798, USA

[5]Department of Physics, University of South Florida, Tampa, FL 33620, USA

**Abstract**

Band gap tuning in two-dimensional transitional metal dichalcogenides (TMDs) is crucial in fabricating new optoelectronic devices. High resolution photoluminescence (PL) microscopy is needed for accurate band gap characterization. We performed tip-enhanced photoluminescence (TEPL) measurements of monolayer MoSe$_2$ with nanoscale spatial resolution, providing an improved characterization of the band gap correlated with the topography compared with the conventional far field spectroscopy. We also observed PL shifts at the edges and investigated the spatial dependence of the TEPL enhancement factors.

# 1. Introduction

The reduced dimensionality of 2D materials leads to distinct properties compared to bulk materials. Unlike graphene, monolayer transitional metal dichalcogenides (TMDs) exhibit direct band gaps and hence attracted worldwide attention for their applications in optoelectronic devices[1], e.g. light emitting diodes[2,3], single photon emitters[4,5], sensing[6,7] and solar cells[8]. Many TMD materials have structures analogous to graphene with covalently bonded $MX_2$ layers which are held together by van der Waals (vdW) interactions, where M is often Mo or S and X is S, Se or Te.

$MoS_2$ is the most widely studied TMD material. Monolayer $MoS_2$ is a direct band gap semiconductor with the band gap at ~ 1.8 eV[9,10]. There are fewer studies carried out on $MoSe_2$. However, $MoSe_2$ may have advantages due to its narrower band gap, narrower line width and tunable excitonic properties[11,12]. Recent studies show that $MoSe_2$ could be a candidate for polaritonics in ambient environment[13,14].

To achieve a more accurate spectroscopic characterization, higher spatial resolution than that provided by the conventional confocal photoluminescence (PL) is needed. Mapping local band gap variations (e.g. exciton/trion population distribution, band structure change due to chemical doping or defects) is essential for fabricating excitonic and valleytronic devices[15,16]. However, confocal microscopy is restricted by the diffraction limit, and therefore is insufficient. Near-field imaging techniques are required to perform nanoscale optical characterization[17–22].

Here we performed first tip-enhanced photoluminescence (TEPL) nanoscale mapping of monolayer $MoSe_2$. We demonstrate higher spatial resolution than the far-field microscopy. We investigated the PL shifts of the edges and spatial dependence of the enhancement factor. Our results may be used to better understand the local properties of 2D materials and to improve optoelectronic devices.

## 2. Results and discussion

AFM topographic image of a cluster of monolayer MoSe$_2$ flakes is shown in Fig.1a. The thickness detected by AFM is ~ 1 nm, which corresponds to a monolayer. The aligned particles (APs) on the edges are similar to those described in the previous reports[23,24,25]. These APs are mixtures of MoO$_x$Se$_y$ generated during the CVD process. Far-field (tip out) and near-field (tip in) intensity maps are shown separately and overlapped with AFM in Figs.1d-1e and 1b-1c, respectively. The intensity was integrated from 785 nm to 835 nm, covering the full PL peak of MoSe$_2$. These maps show no PL signal from the APs. Fig. 1f shows the near-field and far-field PL intensity profiles from the dashed lines A and B shown in Figs. 1d and 1e. Three specific points (I-III and I'-III') were marked on each line for comparison. Clearly, points II and II' distinguish the PL signal from flakes 4-5-6 and flakes 1-2-3 respectively in the near field maps, which are not all visible in the far field maps due to the masking by the APs and due to the lower spatial resolution. Also, the near-field profiles show sharper borders than the far-field profiles. This proves that TEPL can be used to enhance the MoSe$_2$ optical signals and provide higher spatial resolution than the far field spectroscopy.

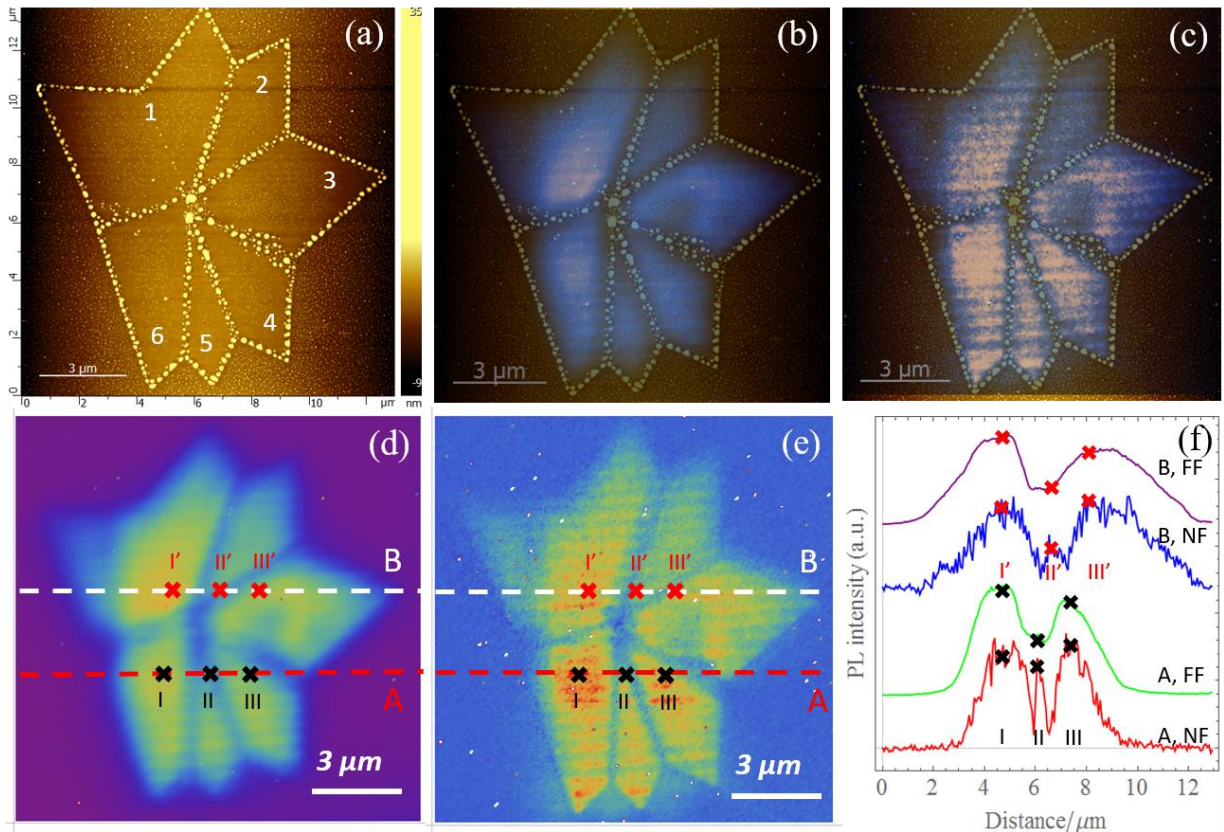

Figure 1. (a) AFM of a cluster of MoSe$_2$ flakes separated by MoO$_x$Se$_y$ aligned particles (APs). (b) AFM overlapped with the far-field photoluminescence (PL) map. (c) AFM overlapped with the near-field PL map. (d) Far-field PL map. (e) Near-field PL map. (f) PL intensity profiles which correspond to the dashed lines A and B shown in (d) and (e). Three points marked I-III and I'-III' on lines A and B, respectively, are used for the comparison between the far-field and near-field signals.

Next, we investigated the spatial dependence of the PL peak position and observed a blue shift around the edges of the CVD-grown monolayer $MoSe_2$. The edges refer to the APs lines shown in Fig. 1a, which are the crystal boundaries of $MoSe_2$. In Figs. 2a and 2b, the blue and red areas represent the integrated intensity in the range of 785-795 nm and 815-820 nm, respectively. Fig. 2c shows the far-field PL peak position map, and the peak position profile of the dashed line C is illustrated in Fig. 2f. Fig. 2g shows the spectra of the points I-V marked in Fig. 2c. Previous work reported edge PL red shift compared to the center in 2D TMDs[24], as well as the blue shifted edge PL[18] and attributed the change of binding energy to the change of M/X ratio (M=Mo, W and X=S, Se)[24] and energy funneling[18], respectively.

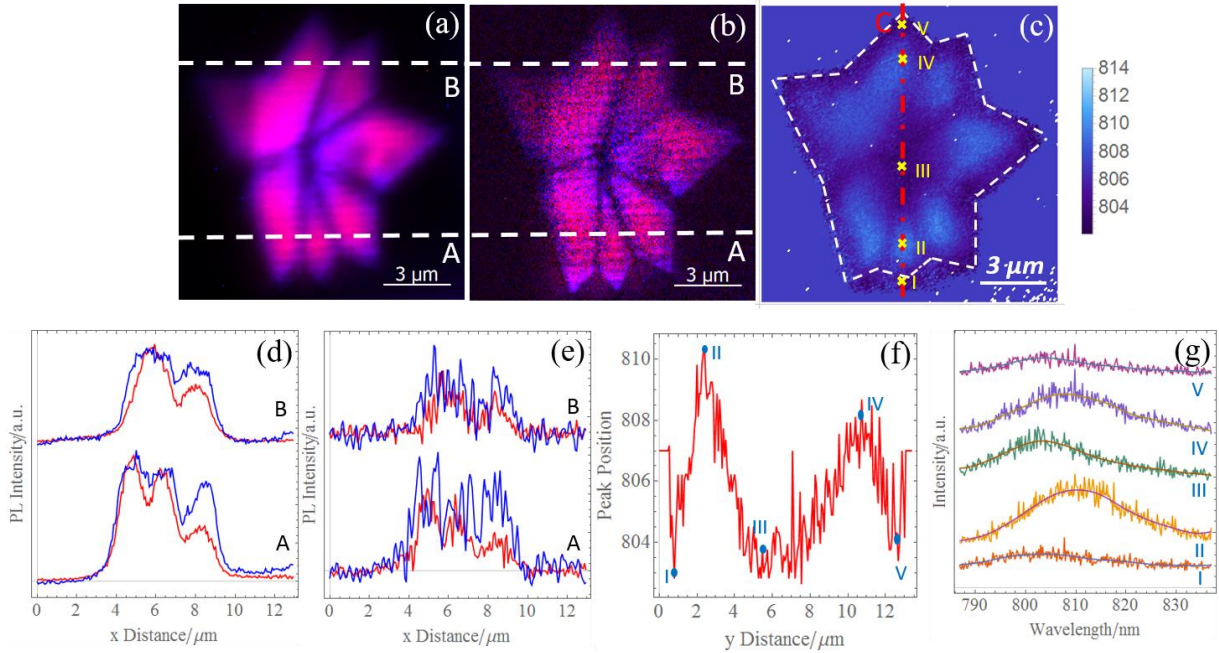

Figure 2. (a) Far-field PL intensity map and (b) near-field PL (TEPL) intensity map. Blue represents integrated intensity from 787 to 795 nm. Red represents integrated intensity from 815 to 820 nm. PL intensity profiles that correspond to dashed lines A, B are shown in (d) and (e). (c) Far-field PL peak position map. Peak position which corresponds to the dashed line C is shown in (f). (g) Far-field spectra from spots I-V marked in (c).

Finally we investigated the spatial dependence of the TEPL enhancemnet factor (EF). Figs. 3a and 3b show the EF maps separately and overlapped with the AFM image respectively. The EF maps were obtained using the standard procedure:

$$EF = \left(\frac{I_{NF}+I_{FF}}{I_{FF}} - 1\right) \times \frac{S_{FF}}{S_{NF}}, \quad S_{FF} = \pi R_{laser}^2, \quad S_{NF} = \pi R_{tip}^2,$$

where the $I_{FF}$ and $I_{NF}$ are the integrated far-field and near-field PL intensities over the full spectral range, and $S_{FF}$ and $S_{NF}$ are the estimated surface areas of the estimated focused far-field and near-field excitation spots, respectively.

The EF maps and profiles B and C in Fig. 3c show that the EF decreased during the tip scan, which was performed from the bottom to the top part of the image. This is attributed to the tip degradation and possible contamination during the TEPL scan. Despite these effects, it is still possible to obtain high resolution imaging of an area of > 100 µm²

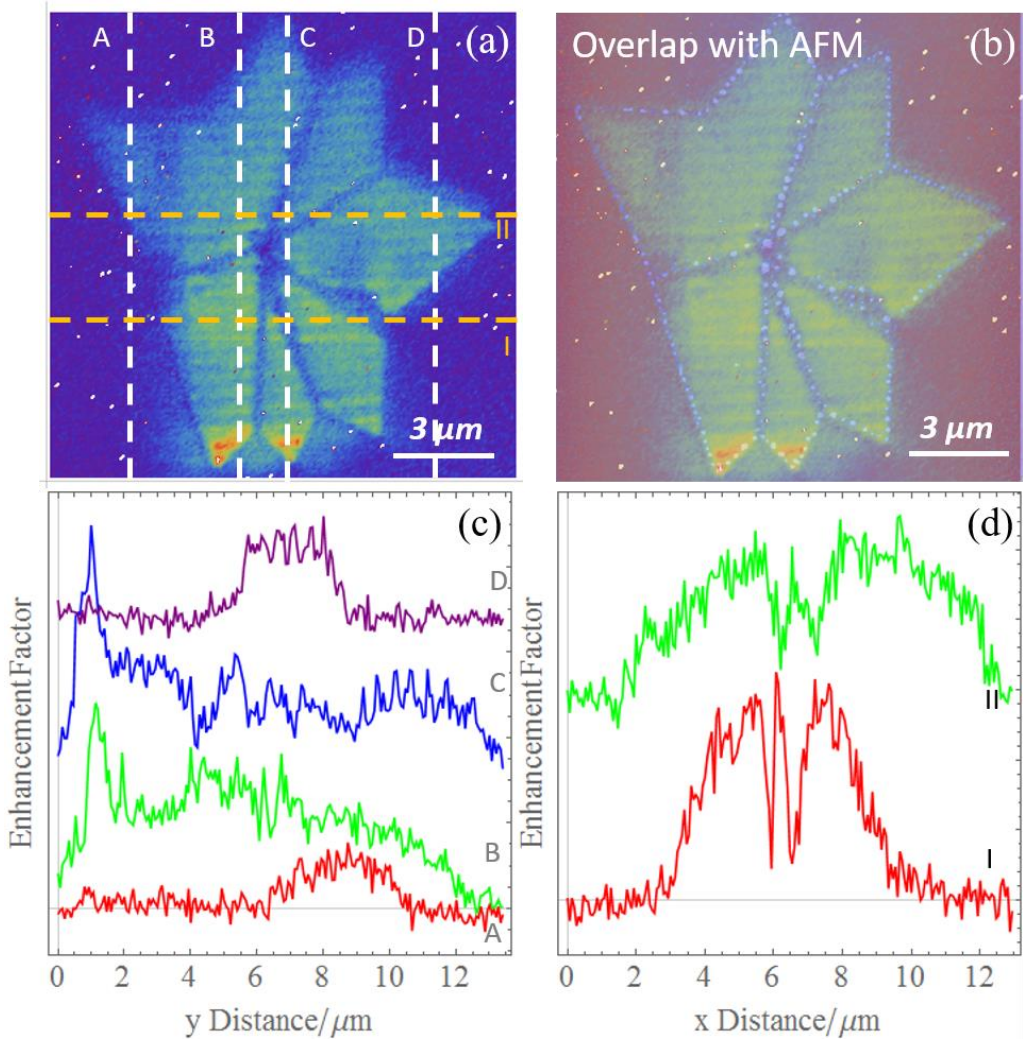

Figure 3. Enhancement factor (EF) map separately (a) and overlapped with AFM (b). EF profiles A, B, C and D (c) and I and II (d) which correspond to the dashed lines in (a).

## 3. Methods

**Synthesis**

Monolayer MoSe$_2$ was synthesized in a home-built CVD furnace. MoO$_3$ and Se powder (Sigma-Aldrich), used as Mo and Se source, were placed at the center and the upstream of the furnace respectively. Clean SiO$_2$/Si substrate faced MoO$_3$ source at the furnace center. Typically, when the furnace temperature ramps to 750 °C, after 15 mins purging, MoO$_3$ will evaporate and react with flowing Se carried by H$_2$ gas (~ 50 sccm), and forms monolayer MoSe2 onto SiO$_2$/Si substrate. After about 10mins growth, the furnace naturally cools down to room temperature.

**Characterization**

Confocal Raman and PL spectroscopies were used to characterize monolayer MoSe$_2$ by Raman peak position and Raman/PL intensity ratio. The experiments were carried out using LabRAM Raman microscope (Horiba).

**Tip-enhanced photoluminescence (TEPL)**

TEPL was carried out using the state-of-the-art commercial system (OmegaScope$^{TM}$ 1000, AIST-NT, coupled with LabRAM Evolution spectrometer, Horiba). Silicon tips were used for AFM and Ag-coated tips with apex radius ~ 20 nm were used for TEPL measurements. The 660 nm laser was focused on the tip apex and the sample was scanned while recording at each point both the near-field and the far-field signals with the tip-sample distance ~ 0.3 nm and ~ 10 nm, respectively. The localized surface plasmons of the tip enhance the electromagnetic field, leading to the enhanced PL emission. The laser power was 1 mW and the step size of the scan was 67 nm, with 0.2 s acquisition time.

## 4. References

1. Wang, Q. H., Kalantar-Zadeh, K., Kis, A., Coleman, J. N. & Strano, M. S. Electronics and optoelectronics of two-dimensional transition metal dichalcogenides. *Nat. Nanotechnol.* **7,** 699–712 (2012).

2. Ross, J. S. *et al.* Electrically tunable excitonic light-emitting diodes based on monolayer WSe2 p-n junctions. *Nat. Nanotechnol.* **9,** 268–272 (2014).

3. Baugher, B. W. H., Churchill, H. O. H., Yang, Y. & Jarillo-Herrero, P. Optoelectronic devices based on electrically tunable p-n diodes in a monolayer dichalcogenide. *Nat. Nanotechnol.* **9,**

262–267 (2014).

4. He, Y.-M. *et al.* Single quantum emitters in monolayer semiconductors. *Nat. Nanotechnol.* **10,** 497–502 (2015).

5. Koperski, M. *et al.* Single photon emitters in exfoliated WSe2 structures. *Nat. Nanotechnol.* **10,** 503–506 (2015).

6. Zuo, X. *et al.* A dual-color fluorescent biosensing platform based on WS2 nanosheet for detection of Hg(2+) and Ag(.). *Biosens. Bioelectron.* **85,** 464–470 (2016).

7. Kuru, C. *et al.* High-performance flexible hydrogen sensor made of $WS_2$ nanosheet-Pd nanoparticle composite film. *Nanotechnology* **27,** 195501 (2016).

8. Pospischil, A., Furchi, M. M. & Mueller, T. Solar-energy conversion and light emission in an atomic monolayer p-n diode. *Nat. Nanotechnol.* **9,** 257–261 (2014).

9. Mak, K. F., Lee, C., Hone, J., Shan, J. & Heinz, T. F. Atomically thin MoS2: A new direct-gap semiconductor. *Phys. Rev. Lett.* **105,** (2010).

10. Splendiani, A. *et al.* Emerging Photoluminescence in Monolayer MoS2. *Nano Lett.* **10,** 1271–1275 (2010).

11. Ross, J. S. *et al.* Electrical control of neutral and charged excitons in a monolayer semiconductor. *Nat. Commun.* **4,** 1474 (2013).

12. Tongay, S. *et al.* Thermally Driven Crossover from Indirect toward Direct Bandgap in 2D Semiconductors: MoSe2 versus MoS2. *Nano Lett.* **12,** 5576–5580 (2012).

13. Lundt, N. *et al.* Monolayered MoSe 2 : a candidate for room temperature polaritonics. *2D Mater.* **4,** 015006 (2017).

14. Kioseoglou, G., Hanbicki, A. T., Currie, M., Friedman, A. L. & Jonker, B. T. Optical polarization and intervalley scattering in single layers of MoS2 and MoSe2. *Sci. Rep.* **6,** 25041

(2016).

15. Baldo, M. & Stojanović, V. Optical switching: Excitonic interconnects. *Nat. Photonics* **3,** 558–560 (2009).

16. Mak, K. F., He, K., Shan, J. & Heinz, T. F. Control of valley polarization in monolayer MoS2 by optical helicity. *Nat. Nanotechnol.* **7,** 494–498 (2012).

17. Bao, W. *et al.* Visualizing nanoscale excitonic relaxation properties of disordered edges and grain boundaries in monolayer molybdenum disulfide. *Nat. Commun.* **6,** 7993 (2015).

18. Park, K.-D. *et al.* Hybrid Tip-Enhanced Nanospectroscopy and Nanoimaging of Monolayer WSe2 with Local Strain Control. *Nano Lett.* **16,** 2621–2627 (2016).

19. Su, W., Kumar, N., Dai, N. & Roy, D. Nanoscale mapping of intrinsic defects in single-layer graphene using tip-enhanced Raman spectroscopy. *Chem. Commun.* **52,** 8227–8230 (2016).

20. Zhang, Y. *et al.* Improving resolution in quantum subnanometre-gap tip-enhanced Raman nanoimaging. *Sci. Rep.* **6,** 25788 (2016).

21. Voronine, D. V., Lu, G., Zhu, D. & Krayev, A. Tip-Enhanced Raman Scattering of MoS2. *IEEE J. Sel. Top. Quantum Electron.* **23,** 138–143 (2017).

22. Lee, Y. *et al.* Near-field spectral mapping of individual exciton complexes of monolayer WS2 correlated with local defects and charge population. *Nanoscale* **9,** 2272–2278 (2017).

23. Li, B. *et al.* Solid–Vapor Reaction Growth of Transition-Metal Dichalcogenide Monolayers. *Angew. Chem. Int. Ed.* **55,** 10656–10661 (2016).

24. Wang, X. *et al.* Chemical Vapor Deposition Growth of Crystalline Monolayer MoSe2. *ACS Nano* **8,** 5125–5131 (2014).

25. Gong, Y. *et al.* Two-Step Growth of Two-Dimensional WSe2/MoSe2 Heterostructures. *Nano Lett.* **15,** 6135–6141 (2015).